\documentclass{moriond}
\usepackage{amsmath}

\def\Journal#1#2#3#4{{#1} {\bf #2}, #3 (#4)}

\def\PLB{{\em Phys. Lett.}  B}
\def\PRL{\em Phys. Rev. Lett.}
\def\PRD{{\em Phys. Rev.} D}
\def\ZPC{{\em Z. Phys.} C}
\def\JHEP{J. High Energy Phys.}

\def\be{\begin{equation}}
\def\ee{\end{equation}}
\def\bea{\begin{eqnarray}}
\def\eea{\end{eqnarray}}

\def\br{\mathcal{B}}

\newcommand{\Photo}{}

\begin{document}
\vspace*{4cm}
\title{$|V_{cs}|$ determination and LFU test in charm decays at BESIII}

\author{P.~L.~Liu~\footnote{email:liupl@ihep.ac.cn}
  \\on behalf of the BESIII collaboration
}
\address{
  Institute of High Energy Physics, Beijing 100049, People's Republic of China 
}

\maketitle\abstracts{In this talk, we present the recent results of charm physics from the BESIII experiment. It covers the studies of the pure leptonic and semi-leptonic decays of charmed hadrons, from which the $|V_{cs}|$ are precisely determined and the LFU is tested. 
}

\section{Introduction}
The BESIII~\cite{bes} experiment at the BEPCII collider started data taking since 2008. For the study of $D$ meson decays, about 7.9\,$\rm{fb}^{-1}$ $e^+e^-$ annihilation data has been collected at center-of-mass energies $E_{\rm{cm}}=3.773$\,\rm{GeV} with the BESIII detector. $\psi(3770)$ dominantly decays into $D\bar{D}$, which provides an ideal place for studying the decays of $D^0$ and $D^+$ mesons. In additon, BESIII has collected an integral luminosity of 7.3\,$\rm{fb}^{-1}$ data at $E_{\rm{cm}}$ between 4.128 to 4.226\,\rm{GeV}. At this region, the $D^{\pm}_{s}$ mesons are produced mainly through the process \mbox{$e^+e^- \rightarrow D^{\star \pm}_{s}D^{\mp}_{s} \rightarrow \gamma(\pi^0) D^{\pm}_{s}D^{\mp}_{s}$}. 
To study the $\Lambda_{c}$ decays, a 4.5\,$\rm{fb}^{-1}$ dataset is collected at $E_{\rm{cm}}$ between 4.600 to 4.699\,\rm{GeV}. At this region, no additional hadrons accompanying the $\Lambda_{c}^+ \bar{\Lambda}_{c}^{-}$ pairs are kinematically allowed.

\section{Determination of $\br(D^+_s\rightarrow l^+\nu_l)$ and $|V_{cs}|$}\label{sec-Ds-lep}
In the standard Model (SM) of particle physics, the $D_s^+$ meson can decay into $l^+\nu_l$ via a virtual $W^+$ boson. The decay rate depends on the $D^+_s$ decay constant $f_{D^+_s}$, in which all of the strong interaction effects between the two initial-state quarks are absorbed.
The decay width of \mbox{$D^+_s\rightarrow l^+\nu_l$} is given by~\cite{dl-lepD}
\begin{equation}
  \Gamma(D^+_s\rightarrow l^+\nu_l)=\frac{G_F^2}{8\pi}|V_{cs}|^2f_{D^+_s}^2m_l^2m_{D_s^{+}}(1-\frac{m_l^2}{m_{D^{+}_s}^2})^2,
\end{equation}
where $G_F$ is the Fermi coupling constant. By measuring the branching fraction of $D^+_s\rightarrow l^+\nu_l$ and taking the $f_{D^+_s}$ from LQCD calculations as input,   $|V_{cs}|$ can be determined to test unitarity of CKM matrix.

To determine the branching fraction of $D^+_s\rightarrow \l^+\nu_\mu$, the tagging method is used. Firstly, $D^-_s$ mesons are reconstructed using 16 hadronic decay modes ($D^-_s\rightarrow K^+ K^- \pi^-$, $K^+ K^- \pi^-\pi^0$ , $\pi^+\pi^-\pi^-$,  $K_S^0 K^-$, $K_S^0 K^-\pi^0$, $K^-\pi^+\pi^-$, $K_S^0 K_S^0\pi^-$, $K_S^0 K^+ \pi^-\pi^-$, $K_S^0 K^- \pi^+\pi^-$, $\eta_{\gamma\gamma} \pi^-$, $\eta_{\pi^+\pi^-\pi^0} \pi^-$, $\eta'_{\pi^+\pi^-\eta_{\gamma \gamma}} \pi^-$, $\eta'_{\gamma\rho^0}\pi^-$, $\eta_{\gamma\gamma} \rho^-$, $\eta_{\pi^+\pi^-\pi^0} \rho^-$, and $\eta_{\gamma\gamma} \pi^+\pi^-\pi^-$), and then $l^+$ is sought out in the remaining tracks. $M_{miss}^2=(E_{beam}-E_{D^-_s}-E_{l^+})^2-(-\overrightarrow{p}_{D^-_s}-\overrightarrow{p}_{l^+})^2$ is used to extract the signal events which peak around~0. Finally, with 7.3\,fb$^{-1}$ of $e^+e^-$ collision data collected at $E_{\rm{cm}}$ from 4.128 to 4.226 \,\rm{GeV}, the branching fraction of $D^+_s\rightarrow \mu^+\nu_\mu$ is measured to be~\cite{PRD108-112001}
\[
  \br(D^+_s\rightarrow \mu^+\nu_\mu)=(5.29\pm0.11\pm0.09)\times10^{-3}.
\]
With combination of $G_{F}$, the mass of $\mu^+$ and $D^+_s$, the lifetime of $D^{+}_s$ and $f_{D^{+}_s}$ from LQCD calculation in the SM, $|V_{cs}|$ is determined to be ~\cite{PRD108-112001}
\[
  |V_{cs}|=0.968\pm0.010\pm0.009.
\]

\begin{figure}[h]
  \centering
  \includegraphics[height=10.0cm,width=0.9\linewidth]{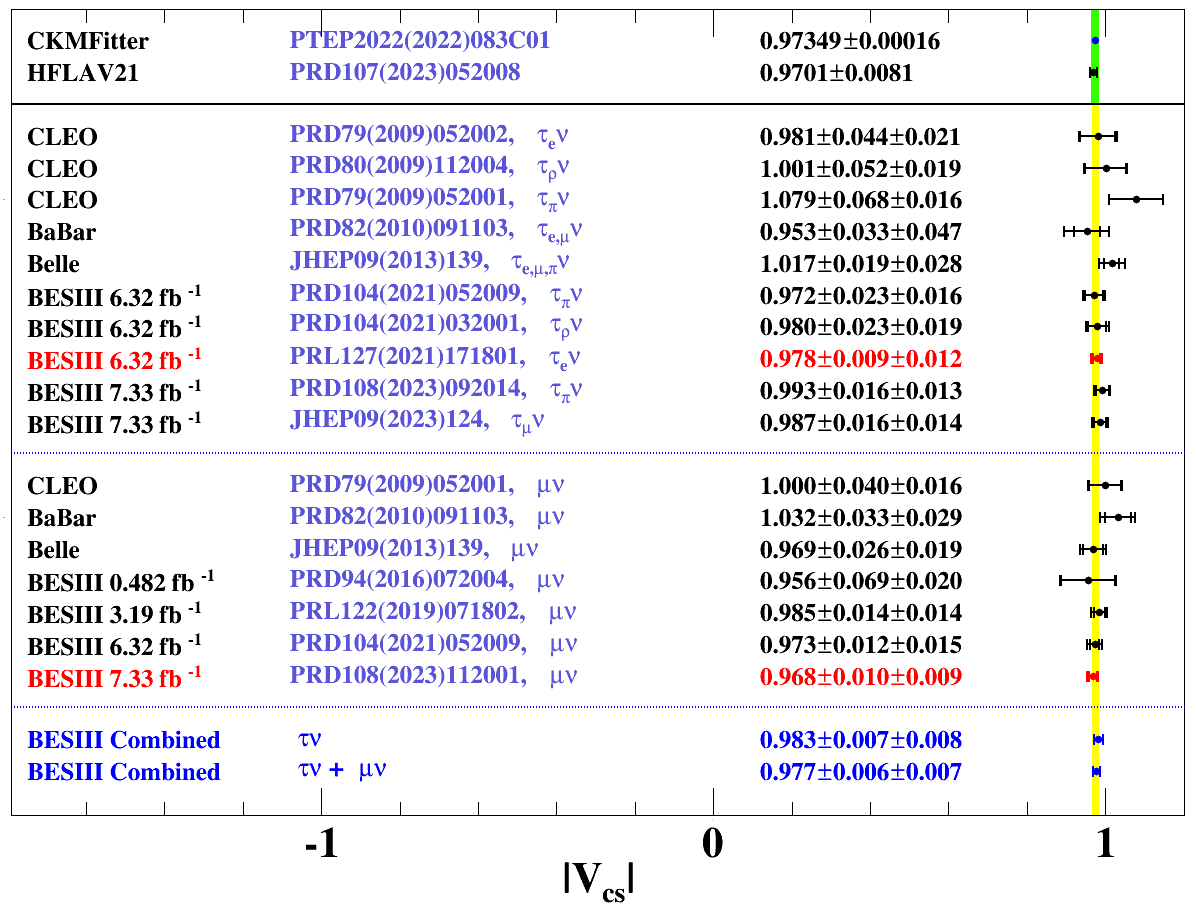}
  \caption{$|V_{cs}|$ determined via $D^+_s\rightarrow l^+\nu_l$. For experimental results, the inner error bar is the statistical uncertainty and the outer is the combined statistical and systematic uncertainty. The green band denotes the CKM Fitter average and the yellow one denotes the experimental average. The last two lines are the BEIII combined results.} 
  \label{fig:Ds-lnu-Vcs}
\end{figure}

For $D^+_s\rightarrow \tau^+\nu_{\tau}$, BESIII reported the measurements of its branching fraction via \mbox{$\tau^+ \rightarrow \rho^+ \bar{\nu}_{\tau}$~\cite{PRD104-032001}}, $\tau^+ \rightarrow \pi^+ \bar{\nu}_{\tau}$~\cite{PRD104-052009,PRD108-092014}, $\tau^+ \rightarrow e^+ \bar{\nu}_{\tau} \nu_{e}$~\cite{PRL127-171801} and $\tau^+ \rightarrow \mu^+ \bar{\nu}_{\tau} \nu_{\mu}$~\cite{JHEP09-124}.  Figure~\ref{fig:Ds-lnu-Vcs} compares all of experimental measurements of $|V_{cs}|$ and the predictions from SM. By combining all of the $\br(D^+_s\rightarrow l^+\nu_l)$ measurement at the BESIII experiment, $|V_{cs}|$ is determined to be $0.977\pm 0.006\pm 0.007$. 

\section{Measurement of $D^{+/0}_{(s)}\rightarrow P l^+\nu_l$ decay dynamics and determination of $|V_{cs}|$}

The differential decay rate of $D^{+/0}_{(s)} \rightarrow P l^+\nu_l$ is written as
\begin{equation}
  \centering
  \frac{d\Gamma}{dq^2}=\frac{G_F^2 |V_{cs}|^2 }{24\pi^3} \frac{(q^2-m_l^2)^2|\overrightarrow{p}|}{q^4m^2_{D^{+/0}_{(s)}}}[(1+\frac{m^2_l}{2q^2})m^2_{D^{+/0}_{(s)}} |\overrightarrow{p}|^2 |f_+(q^2)|^2 + \frac{3m_l^2}{8q^2} (m^2_{D^{+/0}_{(s)}}-m_P^2)^2  |f_0(q^2)|^2 ] ,
  \label{eq:lep-general}
\end{equation}
where $\overrightarrow{p}$ is the three-momentum of the pseudoscalar meson $P$ in the rest frame of the $D^{+/0}_{(s)}$ meson, and $f_{+,0}(q^2)$ represents the hadronic form factors of the hadronic weak current that depend on the square of the four-momentum transfer $q=p_{D^{+/0}_{(s)}}-p_{P}$. These form factors describe strong interaction effects that can be calculated in LQCD.

The differential decay rate of $D^{+/0}_{(s)} \rightarrow P e^+\nu_e$ is simplified as
\begin{equation}
  \centering
  \frac{d\Gamma}{dq^2}=\frac{G_F^2}{24\pi^3}|V_{cs}|^2|f_+(q^2)|^2 |\overrightarrow{p}|^3.
  \label{eq:lep}
\end{equation}
The $D^{+/0}_{(s)}\rightarrow P l^+\nu_l$ dynamics is studied by dividing the events into various $q^2$ intervals. The partial decay rate in each $q^2$ interval is determined with the same method introduced in Sec.~\ref{sec-Ds-lep}. 
To determine the product $f_{+,0}(0)|V_{cs}|$ and other form factor parameters, we fit the measured partial decay rates using Eq.~\ref{eq:lep} with the parametrization of the form factor $f_{+,0}(q^2)$. The form factor parametrizations have several forms. In general, the {\em single pole model}~\cite{single} is the simplest approach to describe the $q^2$ dependence of the form factor. The {\em modified pole model}~\cite{modified} is also widely used in LQCD calculations and experimental studies. The most general parametrization of the form factor is the {\em series expansion}~\cite{series}, which has been shown to be consistent with constraints from QCD. 

\begin{figure}[h]
  \centering
  \includegraphics[height=3.0cm,width=0.4\linewidth]{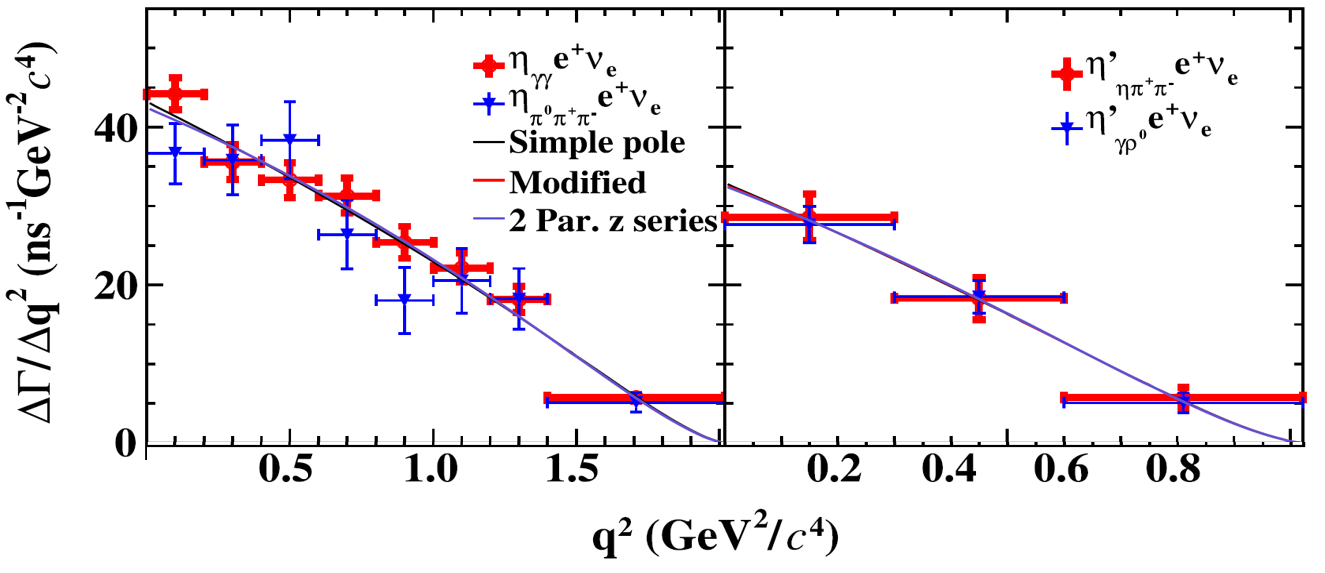}
  \includegraphics[height=2.8cm,width=0.55\linewidth]{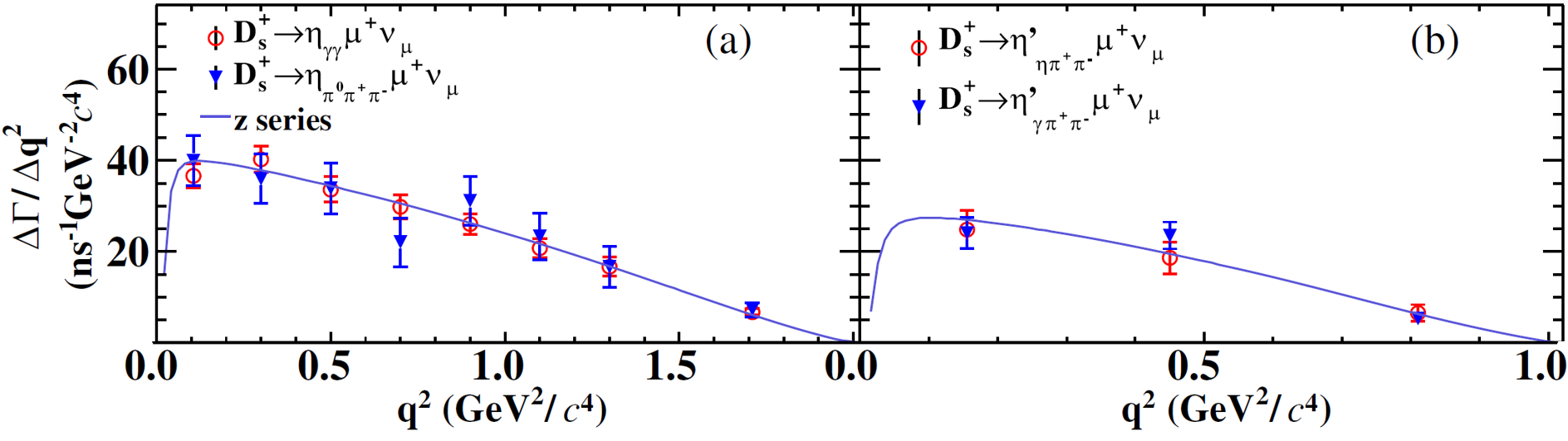}
  \caption{Fits to the differential decays rates of $D^+_s\rightarrow \eta^{('^)} e^+ \nu_e$ (left) and $D^+_s\rightarrow \eta^{(')} \mu^+ \nu_{\mu}$ (right).}
  \label{fig:Ds-eta-l-nu}
\end{figure}

\begin{figure}[h]
  \centering
  \includegraphics[height=8.0cm,width=0.9\linewidth]{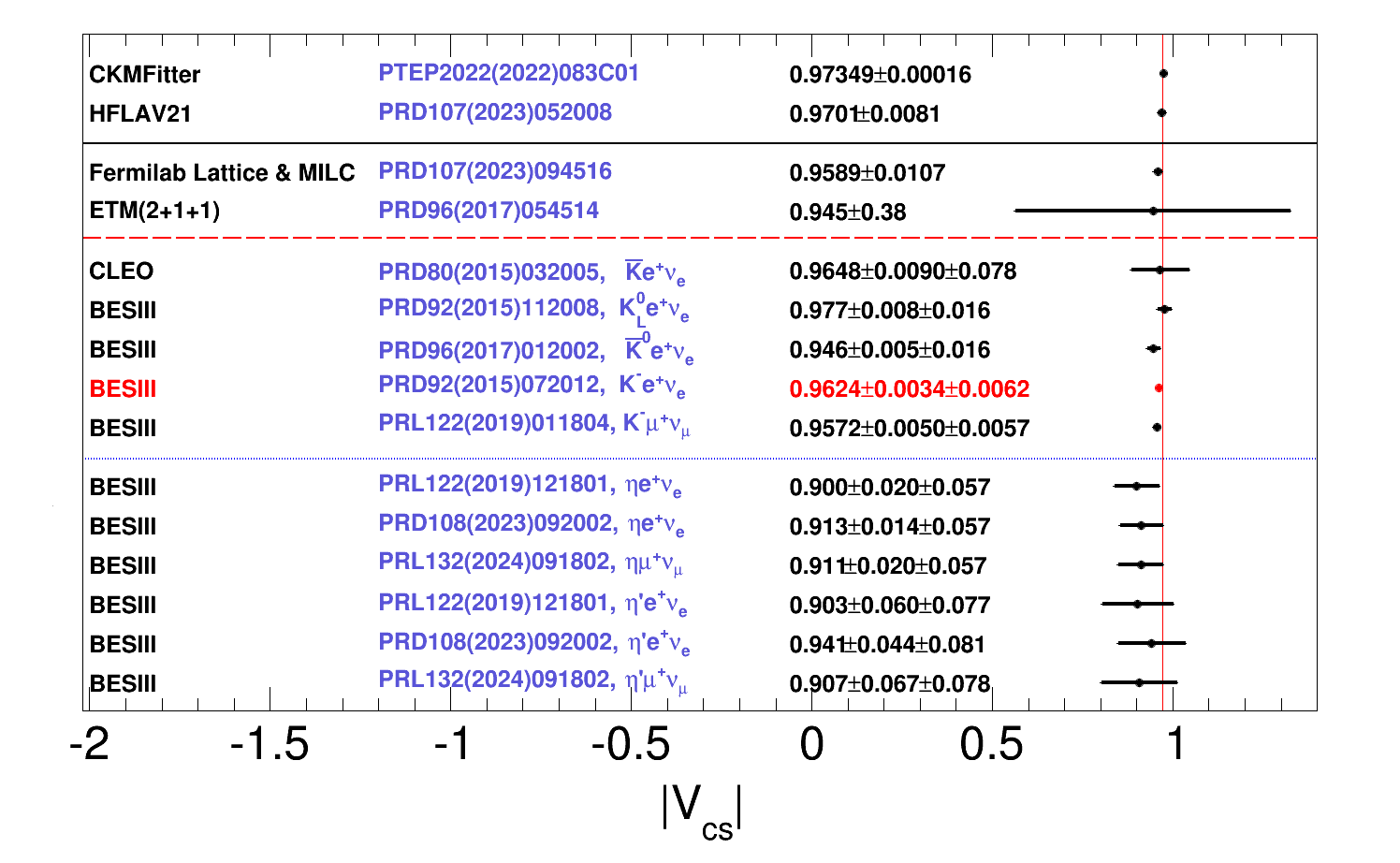}
  \caption{$|V_{cs}|$ determined via $D^{+/0}_{(s)}\rightarrow P l^+\nu_l$.}
  \label{fig:Vcs_semilep}
\end{figure}

BESIII studied the dynamics of $D^0\rightarrow K^- e^+ \nu_e$~\cite{PRD92-072012}, $D^+\rightarrow \bar{K}^0 e^+ \nu_e$~\cite{PRD96-012002}, $D^+\rightarrow K^0_L e^+ \nu_e$~\cite{PRD92-112008}, and $D^0\rightarrow K^- \mu^+ \nu_{\mu}$~\cite{PRL122-011804}. Recently BESIII performed precision measurements of $D^+_s\rightarrow \eta^{(')} e^+ \nu_e$~\cite{PRD108-092003} and $D^+_s\rightarrow \eta^{(')} \mu^+ \nu_{\mu}$~\cite{PRL132-091802}. The dependence on $q^2$ of the differential decay rates are shown in Figure~\ref{fig:Ds-eta-l-nu}. By combining the $f_{+,0}(0)$ given by theoretical calculations with $f_{+,0}(0)|V_{cs}|$ extracted from the series expansion, $|V_{cs}|$ determined with $D^{+/0}_{(s)}\rightarrow P l^+\nu_l$ are listed in Figure~\ref{fig:Vcs_semilep}.

\section{Study of the semileptonic decay $\Lambda_{c}^+\rightarrow \Lambda e^+ \nu_e$}
The differential decay rate of $\Lambda_{c}^+\rightarrow \Lambda e^+ \nu_e$ can be expressed in terms of $q^2$ by 
\begin{equation}
  \centering
  \frac{d\Gamma}{dq^2}=\frac{G_F^2 |V_{cs}|^2}{192\pi^3 m_{\Lambda_{c}}^2}\times P q^2 \times [|H_{\frac{1}{2} 1}|^2 + |H_{-\frac{1}{2} -1}|^2 + |H_{\frac{1}{2} 0}|^2 + |H_{-\frac{1}{2} 0}|^2 ],
  \label{eq:lep-lambda}
\end{equation}
where $H_{\lambda_{\Lambda}\lambda_{W}}= H^V_{\lambda_{\Lambda}\lambda_{W}} - H^A_{\lambda_{\Lambda}\lambda_{W}}$ and $H^{V(A)}_{-\lambda_{\Lambda}-\lambda_{W}}=+(-)H^{V(A)}_{\lambda_{\Lambda}\lambda_{W}}$.  $P$ is the magnitude of the $\Lambda$ momentum in the $\Lambda_{c}$ rest frame.  
The helicity amplitudes are related to four form factors by
\begin{equation}
 \begin{aligned}
   &H_{\frac{1}{2} 1}^V = \sqrt{2Q_-}f_{\perp}(q^2) , ~~~~~~ H_{\frac{1}{2} 1}^A = \sqrt{2Q_+}g_{\perp}(q^2),  \\
   &H_{\frac{1}{2} 0}^V = \sqrt{Q_-/q^2}f_+(q^2)(m_{\Lambda_{c}}+m_{\Lambda})  , ~~~~~~ H_{\frac{1}{2} 0}^A = \sqrt{Q_+/q^2}g_+(q^2)(m_{\Lambda_{c}}-m_{\Lambda}) , \\ 
 \end{aligned}
  \label{eq:helicity-FF}
\end{equation}
where $Q_{\pm}=(m_{\Lambda_{c}}\pm m_{\Lambda})^2-q^2$. The form factors $f_{\perp,+}(q^2)$ and $g_{\perp,+}(q^2)$ are defined following a $z$-expansion of the parameters as in Ref.~\cite{PRL129-231803}.

\begin{figure}[h]
  \centering
  \includegraphics[height=5.0cm,width=0.5\linewidth]{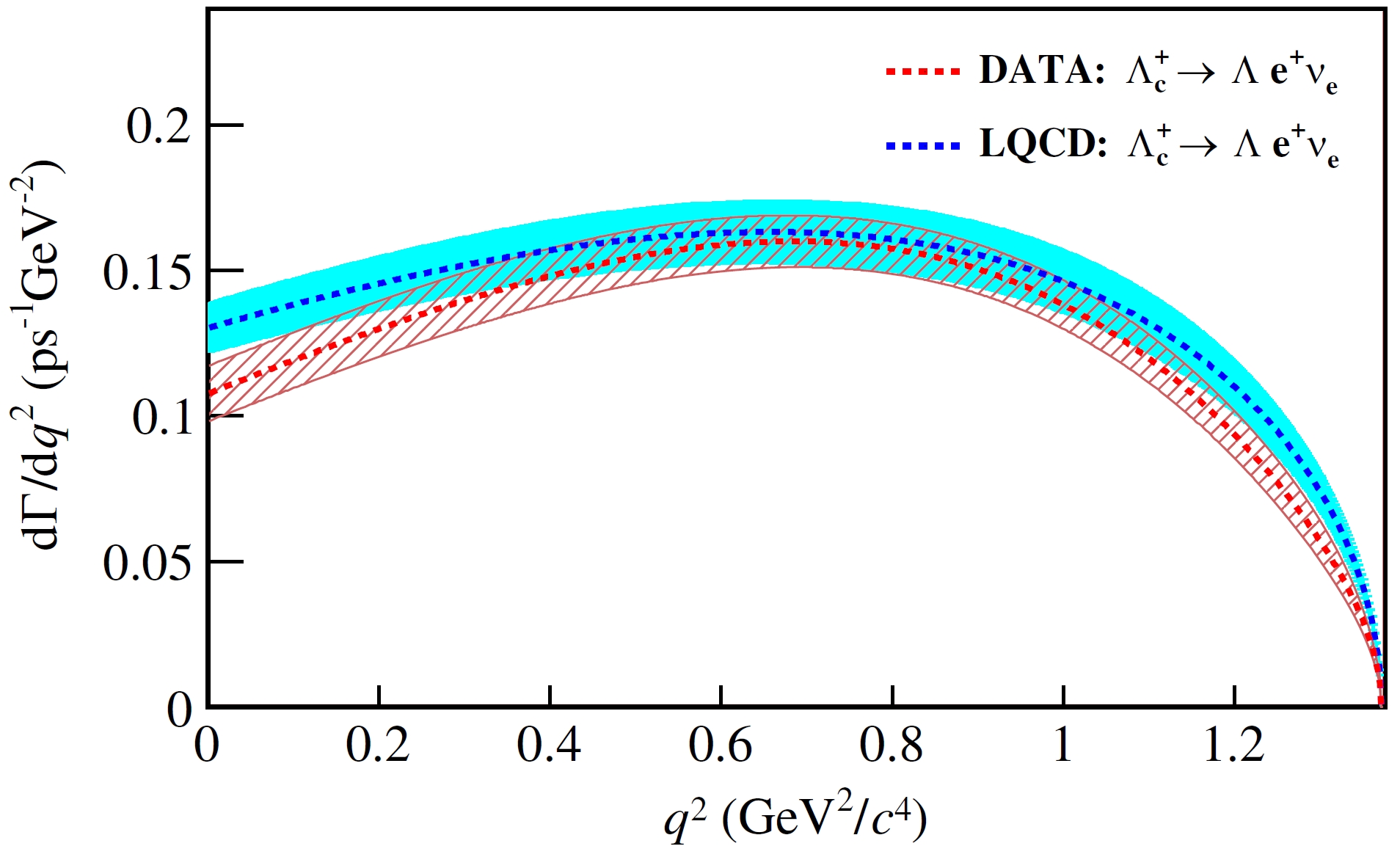}
  \caption{comparison of the differential decay rates with LQCD predictions.}
  \label{fig:Lambda-c-dGamma}
\end{figure}

Figure~\ref{fig:Lambda-c-dGamma} shows comparison of the differential decay rate between measurements in Ref.~\cite{PRL129-231803} and in LQCD calculation, which gives fair agrement throughout the $q^2$ region. This is the first direct comparison on the differential decay rate with that obtained from LQCD calculation.   

The branching fraction is measured to be $\br(\Lambda_{c}\rightarrow \Lambda e^+ \nu_e)=(3.56\pm0.11\pm0.07)\%$, which is the most precise measurement to date, with the same method introduced in Sec.~\ref{sec-Ds-lep}. Combining the measured $\br(\Lambda_{c}\rightarrow \Lambda e^+ \nu_e)$ and the $q^2$-integrated rate predicted by LQCD, we determine $|V_{cs}|=0.936\pm0.030$, which is consistent with the values obtained via $D^{+/0}_{(s)}\rightarrow P l^+\nu_l$.  

\section{LFU test}

Lepton flavor universality (LFU) is usually thought of as a basic property of the SM. It postulates that the coupling between the three families of leptons and gauge bosons do not depend on the lepton flavor.  Experimental studies of (semi-)leptonic decays of charmed hadrons are important to test LFU and explore possible new physics. 

\subsection{LFU test in  $D_s^+ \rightarrow  l^+ \nu_l$}
\begin{figure}[h]
  \centering
  \includegraphics[height=6.5cm,width=0.8\linewidth]{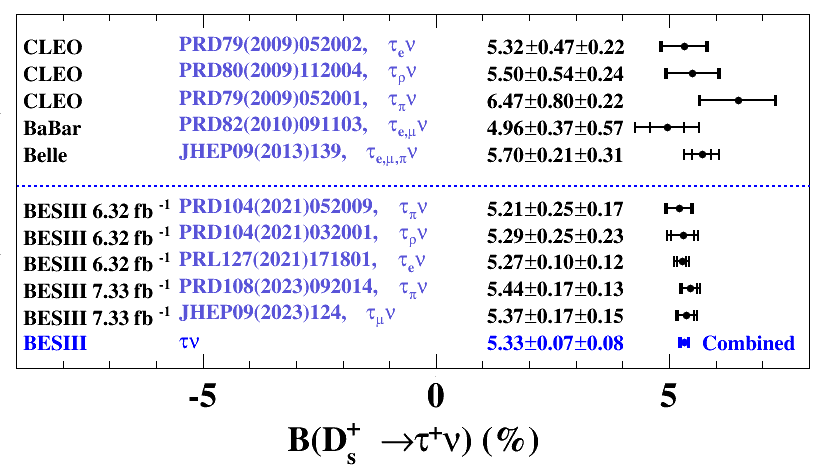}
  \caption{Comparison of the branching fractions experimentally measured, where the inner error bar is the statistical uncertainty and the outer is the combined statistical and systematic uncertainty. The last line is the BESIII combined result.} 
  \label{fig:br-Ds-tau-nu}
\end{figure}

In Figure~\ref{fig:br-Ds-tau-nu}, the branching fraction measurements of $D_s^+ \rightarrow \tau^+ \nu_\tau$ using the decays $\tau^+ \rightarrow \rho^+ \bar{\nu}_{\tau}$~\cite{PRD104-032001}, $\tau^+ \rightarrow \pi^+ \bar{\nu}_{\tau}$~\cite{PRD104-052009,PRD108-092014}, $\tau^+ \rightarrow e^+ \bar{\nu}_{\tau} \nu_{e}$~\cite{PRL127-171801} and $\tau^+ \rightarrow \mu^+ \bar{\nu}_{\tau} \nu_{\mu}$~\cite{JHEP09-124} at the BESIII experiment are  compared with those from other experiments. We determine the average BF measured at the BESIII experiment tobe $\br(D_s^+ \rightarrow \tau^+ \nu_\tau)=(5.33\pm0.07\pm0.08)\%$.  

Using the BESIII measured value of $\br(D_s^+ \rightarrow \mu^+ \nu_\mu)=(5.29\pm0.11\pm0.09)\times 10^{-3}$, we obtain the ratio of branching fractions $R_{\tau/\mu}=10.05\pm0.35$, which agrees with the SM prediction 9.75.  

\subsection{LFU test in $D^{+/0}_{(s)}\rightarrow P l^+\nu_l$  and  $\Lambda_{c}\rightarrow \Lambda l^+ \nu_l$}
The ratios of branching fractions $R_{\mu/e}$ for $D^+ \rightarrow \eta l^+ \nu_l$, $D^+ \rightarrow \omega l^+ \nu_l$, and $D^+ \rightarrow \rho l^+ \nu_l$, which are determined by using the branching fractions measured at the BESIII experiment, are listed in Table~\ref{table:ratio-D+}. There is no significant deviaton from the standard model predictions. 
\begin{table}[h]
  \centering
  \caption{LFU test in semileptonic decays of $D^+$ by measuring branching fractions}
  \begin{tabular}{llcc}
    \hline
    mode        & Ref. & $R_{\mu/e}$ & SM prediction \\
    \hline
    $D^+\rightarrow \eta l^+\nu_l$       &   PRL124(2020)231801      &    0.91$\pm$0.13 &   0.97-1.00 \\
    $D^+\rightarrow \omega l^+\nu_l$     &   PRD101(2020)072005      &    1.05$\pm$0.14 &   0.93-0.99 \\
    $D^+\rightarrow \rho l^+\nu_l$       &   PRD104(2021)L091103     &    0.90$\pm$0.11 &   0.93-0.96 \\
    \hline
  \end{tabular}
  \label{table:ratio-D+}
\end{table}

\begin{figure}[h]
    \centering
      \includegraphics[height=2.5cm,width=0.6\linewidth]{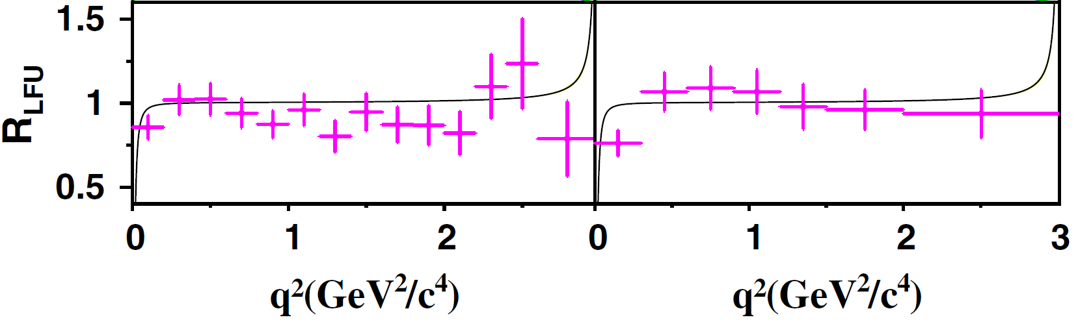}
      \includegraphics[height=2.6cm,width=0.3\linewidth]{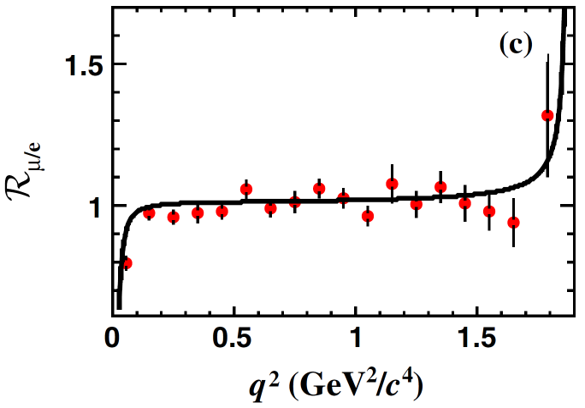}
      \includegraphics[height=2.5cm,width=0.3\linewidth]{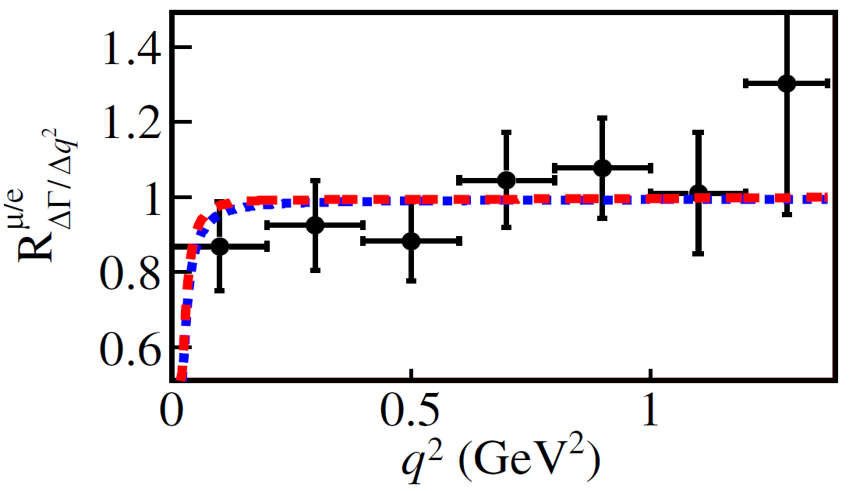}
      \includegraphics[height=2.6cm,width=0.6\linewidth]{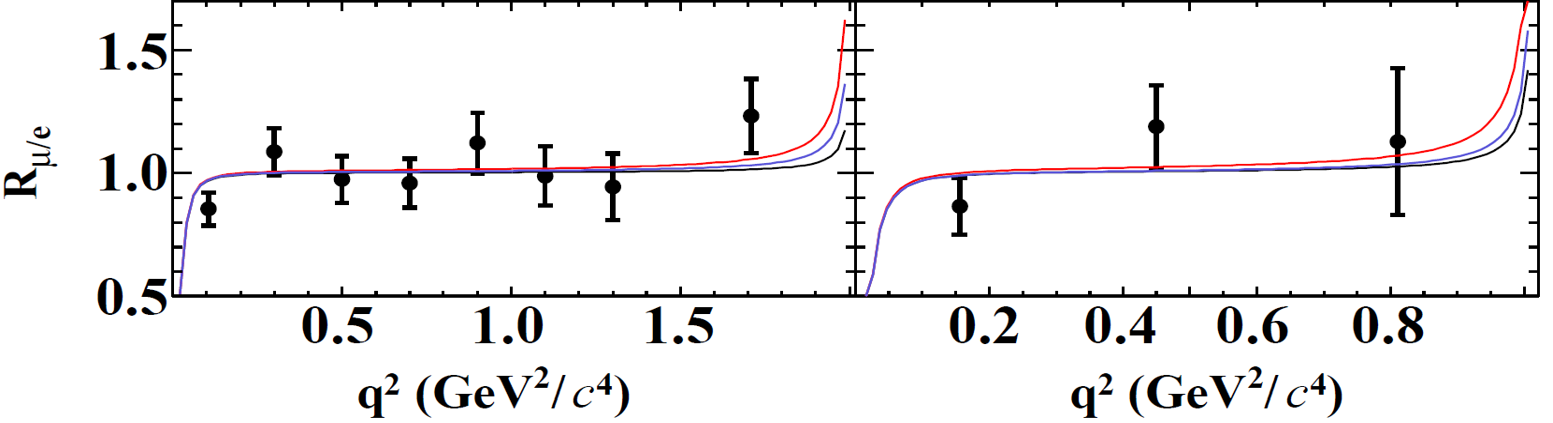}
\caption {The measured $R_{\mu/e}$ combining the two signal channels in each $q^2$ interval. The dots with error bars are data, where statistical and systematic uncertainties are both included. The curves show those predicted by SM.}
\label{fig:R-semi-q2}
\end{figure}

In addition, the raios of the decay widths are examined in different $q^2$ intervals in the decays $D^{0/+} \rightarrow \pi^{-/0} l^+ \nu_l$~\cite{PRL121-171803}, $D^{0} \rightarrow K^{-} l^+ \nu_l$~\cite{PRL122-011804}, $\Lambda_{c}^+\rightarrow \Lambda l^+ \nu_L$~\cite{PRD108-L031105}, and $D^+_s\rightarrow \eta^{(')} \l^+ \nu_{l}$~\cite{PRL132-091802} (left to right and top to bottom in Figure~\ref{fig:R-semi-q2}), by analyzing the dynamics of these decays. Below $q^2=0.1\rm{GeV}^2/c^4$, $R$ is significantly lower than 1 due to smaller phase space for the $l=\mu$ channel with nonzero muon mass. Above $q^2=0.1\rm{GeV}^2/c^4$, $R$ is close to 1. They are also consistent with the SM predictions.

\section{Summary}
Based on the data samples collected at $D\bar{D} / D_s D_s^{\star} / \Lambda_{c}^+ \bar{\Lambda}_{c}^{-}$ mass threshold, (semi-)leptonic decays of charmed hadrons have been studied at the BESIII experiment. The measurement of the CKM matrix element $|V_{cs}|$ is at 1\% precision level. No evidence of LFU violation is found at 1.5\% precision level. BESIII just finished taking the data at 3.773\,\rm{GeV}, and collected 20\,$\rm{fb}^{-1}$ of $D\bar{D}$ sample. A combined study of $D^{0/+}\rightarrow K^{-/0} e^+ \nu_e$ and $D^{0/+}\rightarrow K^{-/0} \mu^+ \nu_{\mu}$ will provide a higher precise measurement of $|V_{cs}|$. The LFU test will be performed with a higher precision.


\end{document}